\documentclass[conference]{IEEEtran}
\IEEEoverridecommandlockouts
\usepackage{cite}
\usepackage{amsmath,amssymb,amsfonts}
\usepackage{algorithmic}
\usepackage{graphicx}
\usepackage{textcomp}
\usepackage{xcolor}
\usepackage[colorlinks,urlcolor=blue,linkcolor=blue,citecolor=blue]{hyperref}
\usepackage{xspace}
\usepackage{color} 
\usepackage{setspace}
\usepackage{url}
\usepackage[colorlinks,citecolor=blue]{hyperref}
\usepackage{booktabs}
\usepackage{tabularx}
\usepackage{todonotes}
\usepackage{tikz}

\newcommand{\flashx}{Flash-X\xspace}
\newcommand{\flash}{FLASH\xspace}
\newcommand{\flashtest}{FlashTest\xspace}

\newcommand{\setuptool}{setuptool\xspace}

\def\BibTeX{{\rm B\kern-.05em{\sc i\kern-.025em b}\kern-.08em T\kern-.1667em\lower.7ex\hbox{E}\kern-.125emX}}
    
\begin{document}
\title{Framework and Methodology for Verification of a Complex Scientific Simulation Software, Flash-X
\thanks{US DOE ASCR}}

\author{\IEEEauthorblockN{1\textsuperscript{st} Akash Dhruv}
\IEEEauthorblockA{\textit{Mathematics and Computer Science} \\
\textit{Argonne National Laboratory}\\
Lemont, IL USA \\
adhruv@anl.gov}
\and
\IEEEauthorblockN{2\textsuperscript{nd} Rajeev Jain}
\IEEEauthorblockA{\textit{Mathematics and Computer Science} \\
\textit{Argonne National Laboratory}\\
Lemont, IL USA\\
rajeeja@mcs.anl.gov}
\and
\IEEEauthorblockN{3\textsuperscript{rd} Jared O'Neal}
\IEEEauthorblockA{\textit{Mathematics and Computer Science} \\
\textit{Argonne National Laboratory}\\
Lemont, IL USA \\
joneal@anl.gov}
\and
\hspace{3.5cm} \IEEEauthorblockN{4\textsuperscript{th} Klaus Weide}
\IEEEauthorblockA{\textit{\hspace{3.5cm} Department of Computer Science} \\
\hspace{3.5cm} \textit{University of Chicago}\\
\hspace{3.5cm} Chicago, IL USA \\
\hspace{3.5cm} kweide@uchicago.edu}
\and
\IEEEauthorblockN{5\textsuperscript{th} Anshu Dubey}
\IEEEauthorblockA{\textit{Mathematics and Computer Science} \\
\textit{Argonne National Laboratory}\\
Lemont, IL USA \\
adubey@anl.gov}
}

\maketitle

\begin{abstract}
Computational science relies on scientific software as its primary
instrument for scientific discovery. Therefore, similar to the use of other types
of scientific instruments, correct software and the correct operation of
the software is necessary for executing
rigorous scientific investigations. Scientific software verification can be
especially difficult, as users typically need to modify the software as part
of a scientific study.
Systematic methodologies for building test suites for
scientific software are rare in the literature. Here, we describe a
methodology that we have developed for \flashx, a community simulation
software for multiple scientific domains, that has composable
components that can be permuted and combined in a multitude of ways to
generate a wide range of applications. Ensuring sufficient code coverage by a
test suite is
particularly challenging due to this composability. Our methodology includes
a consideration of trade-offs between meeting software quality goals,
developer productivity, and meeting the scientific goals of the Flash-X user
community.
\end{abstract}

\begin{IEEEkeywords}
Verification, Methodology, Test suite, Scientific software, Multiphysics
\end{IEEEkeywords}

\section{Introduction}
Scientific processes continue to rely on software as an important tool for data acquisition, analysis, and discovery. Therefore, similar to the use of other types of scientific instruments, correct software and the correct operation of the software is necessary for executing rigorous scientific investigations. Although the lifecycles of software and hardware instruments differ significantly, both need rigorous maintenance and upgrades to remain useful. In particular, software upgrades are often incremental in nature and require seamless operation of testing and verification frameworks to allow flexibility while avoiding degradation.

\flashx  \cite{dubey2022flash} is a multiscale, multiphysics scientific software
package designed to simulate physical systems that are
modeled using a combination of partial differential equations (PDEs),
ordinary differential equations (ODEs), and a few other motifs
(e.g., algebraic or tabulated equations of state, \textit{etc}.). It is built with
composable components that can be permuted and combined in a multitude of
ways to generate a wide range of applications for domains such as
astrophysics\cite{harris2022exascale} and computational fluid dynamics
\cite{chawdhary2018immersed}. Since the
code is often being developed at the same time as it is being used for
executing scientific simulations, ensuring correct functionality requires extensive
regular testing that accounts for various modes of use with different
interoperability constraints for different applications. \flashx is
not unique in this regard. Other multiphysics applications with
varying degrees of composability face similar challenges. For \flashx
we have developed a framework and a methodology with accompanying
tools that we believe will be useful to other applications
facing similar challenges.

\section{\flashx and Testing Requirements}
While \flashx has been described in detail in \cite{dubey2022flash}, we provide
here a brief overview for completeness. The primary functionality of the code
is to solve the Euler or Navier-Stokes equations for compressible and
incompressible flows on a discretized mesh.
The resolution of the mesh can be fixed to a uniform grid across the
domain (UG) or be dynamically controlled with
adaptive mesh refinement (AMR)\cite{bergerColella:1989}, where different
parts of the domain are potentially assigned different resolutions based on the
evolving characteristics of the local physical solution.
Source terms and equations of state can be added
to either formulation. Several source terms, such as nuclear burning \cite{Timmes1999},
are included in the distribution, and several others are expected to
be added by the user community depending on their needs.  The code
includes solvers for hyperbolic, parabolic, and elliptic PDEs; ODE
solvers for certain equations of state (EOS) and burning networks; as well as
algebraic solvers for both EOS and source terms. A Lagrangian
framework is built into the code that can manage particles in purely passive
mode as tracers, or in active mode for particle-in-cell or N-body
simulations. The code interfaces with math libraries such as PETSc
\cite{petsc-user-ref} and Hypre \cite{falgout2000} to solve linear systems if needed.

An important characteristic of \flashx is that it does not build an
executable out of the whole code. With the help of the configuration layer and
assisted by code translation tools \cite{Dubey2009,RudiONeilWahibEtAl2021,DubeyPPAM2022},
each application assembles its
desired code components. Therefore, each instance of an application
compiles only a fraction of the code. A code component can be
a code snippet from a function, a whole function, or a collection of
functions, and it can have alternative implementations that can each serve a different purpose. Some are meant for
different physical regimes or different fidelity requirements, while
others are meant to account for different hardware architectures. 
Components can be self-describing by encoding metainformation
about how they are to be used in the assembled application. The choice of
the alternative implementation is determined by the
configuration layer based upon the encoded metainformation and
command-line instructions.

An application instance in \flashx resides in a specialized code unit, {\em Simulation}, with a subunit {\em SimulationMain} containing different implementations for each specific problem represented by {\em $<$simulation-name$>$}, such as {\em Sod}, {\em SNIa Double Detonation} or {\em Pool Boiling}. Each simulation is said to be an
alternative implementation of the unit's API configured using components from source code, initial conditions, composition in terms of additional species (to support nuclear or
chemical reaction networks), and any needed customizations (see \cite{flashx-userguide}
for more details). These
components also become the basis for the testing infrastructure. Unit
tests are cast as their own {\em $<$simulation-name$>$} with minimal
dependencies.  

\begin{figure}[h]
\begin{center}
\includegraphics[width=0.375\textwidth]{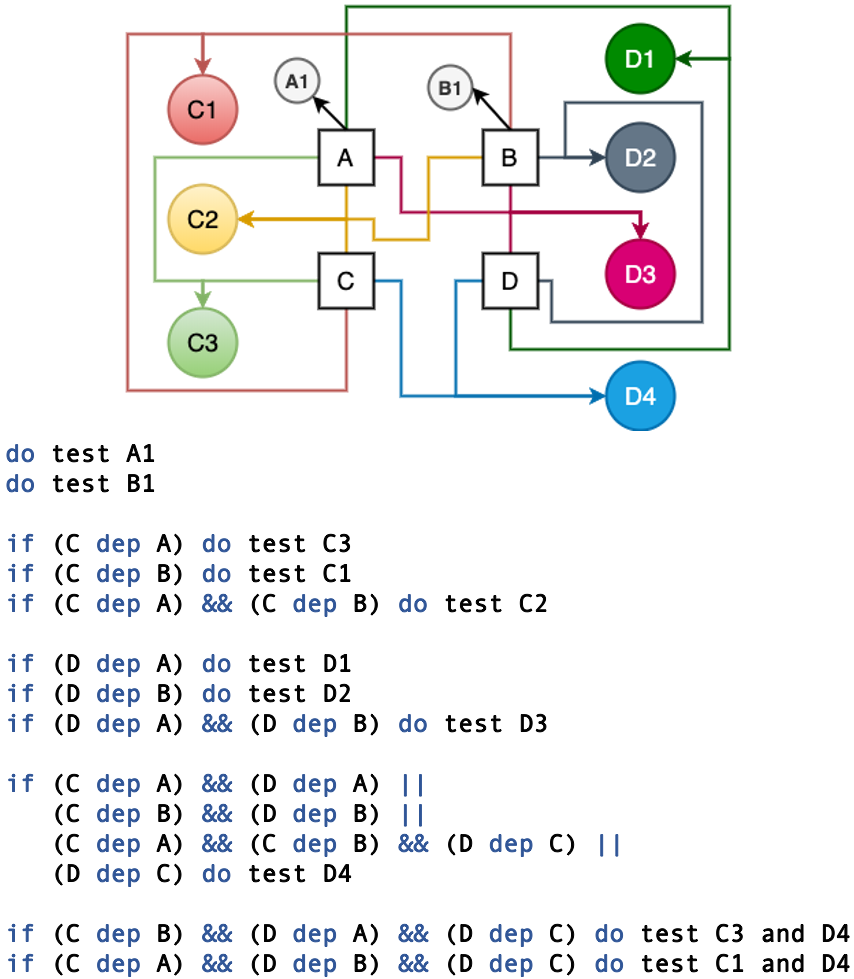}
\caption{Providing unit-test-like coverage through scaffolding for code sections that are
  not easily amenable to unit testing. Code components, {\bf A}, {\bf B}, {\bf C} and {\bf D} can be tested
  using different combinations of tests, {\bf A1}, {\bf B1}, {\bf C1-C3} and {\bf D1-D4} based on their mutual inter-dependence. The code snippet covers conditions and possible combinations of the scaffolding.}
\label{fig:scaffolding}
\end{center}
\end{figure}

Few ``pure'' unit tests exist in the \flashx
test suite. Instead of testing each code component individually, the test suite is formulated in the
form of scaffolding as explained in Figure \ref{fig:scaffolding}. Here,
if components {\bf A} and {\bf B} have unit tests {\bf A1} and {\bf B1},
respectively, then a code unit {\bf C}, which depends on {\bf A} or
{\bf B} or a combination of the two, can be tested using one of the
tests {\bf C1-C3} as outlined in the accompanying code snippet in the figure. A similar strategy
is applied for the tests {\bf D1-D3} for a separate code unit {\bf D}. However, if {\bf D} depends on {\bf C} and {\bf C} depends on {\bf A} and {\bf B}, simply performing the test {\bf D4} along with {\bf A1}, {\bf B1} and {\bf C2} will suffice. In this case {\bf C2} provides a scaffold for testing {\bf D}.

For context, components {\bf A} and {\bf
    B} may represent a stenciled computation for the advection and diffusion terms of a PDE. These components are placed under the code unit {\em
    Stencils}, and are used by the units {\em HeatAD}  ({\bf C}) and {\em
    IncompNS} ({\bf D}) to solve for heat transfer and incompressible
fluid flow, respectively. A typical set of {\bf A1}, {\bf B1} and {\bf C2} in this case
involves testing the scalar transport of temperature through simple advection and diffusion problems. Test {\bf D4} includes testing the transport of heat flux in turbulent fluid flows using a fractional step solution of the incompressible Navier-Stokes equations.
  For this example, the tests involving components {\bf D} are computationally expensive compared to the tests {\bf C1-C3}, and therefore the scaffolding strategy enables the selection of tests to optimize the coverage versus execution time trade-off \cite{dubey2018methodology}.

The regression tests described above can often be generated by
carefully calibrating the runtime parameters of the existing
application instances (see Section~\ref{testcomps}). Typically, if a new component is added that
does not have coverage from existing applications, it is required to
also submit an application setup and a selection of parameters that
can be used in a test configuration of the application in the pull request to be accepted. 

Exhaustive coverage for testing all possible permutations and combinations of components
in \flashx is
neither feasible nor desirable. Some combinations may be
mathematically possible but physically impossible, while the requisite
functionality check for some combinations may be adequately covered by
others through the scaffolding described earlier. In the example above, the success of the test {\bf D4} makes {\bf A1}, {\bf B1} and {\bf C2} redundant. 
However, for a quick diagnosis of the cause of failure,
it is required that we build more tests than strictly necessary. Also,
because the code is not stationary,  it is prudent to
build more tests when a code component is originally included in the
source in case testing needs change due to changes in
interoperability. Therefore, we have two requirements for the testing
framework: (1) it should be possible for the contributors to specify
the configuration and runtime parameters of each relevant application
instance in the code repository as a possible test, and (2) the framework requires a
mechanism to select tests from all those specified in the code repository in
order to construct a particular test suite for execution on a particular
platform with a particular software stack.
In the next section, we describe how these elements of
our testing methodology come together.

\section{Testing Framework}
\flash instituted regular regression testing before any standard
testing frameworks were available for automating the
process. Consequently, \flashtest \cite{dubey2015ongoing} was
developed in-house and served as a testing framework for \flash, the
predecessor of \flashx. We find it to be the best option to run our test suite for \flashx due to similarities in design and organization of the source code between the two.
To inspect the results of the execution of the test suite, the accompanying
web interface, FlashTestView, continues to be the most effective
option. In the following, we describe what constitutes a test in \flashx and how it fits
into the testing framework.

\subsection{Components of a Test} \label{testcomps}
{The {\em $<$simulation-name$>$} configuration of the subunit {\em Simulation/SimulationMain} discussed above also contains a set of files with extension {\em
  .par}, referred to as {\em parfiles}, which provide values of the runtime
parameters related to physics, dimensions, and coordinate system to be used in the application setup. 
An application is likely to
have at least one {\em .par} file for each of the viable configurations it
supports.}

\begin{figure}[h]
\begin{center}
\includegraphics[width=0.48\textwidth]{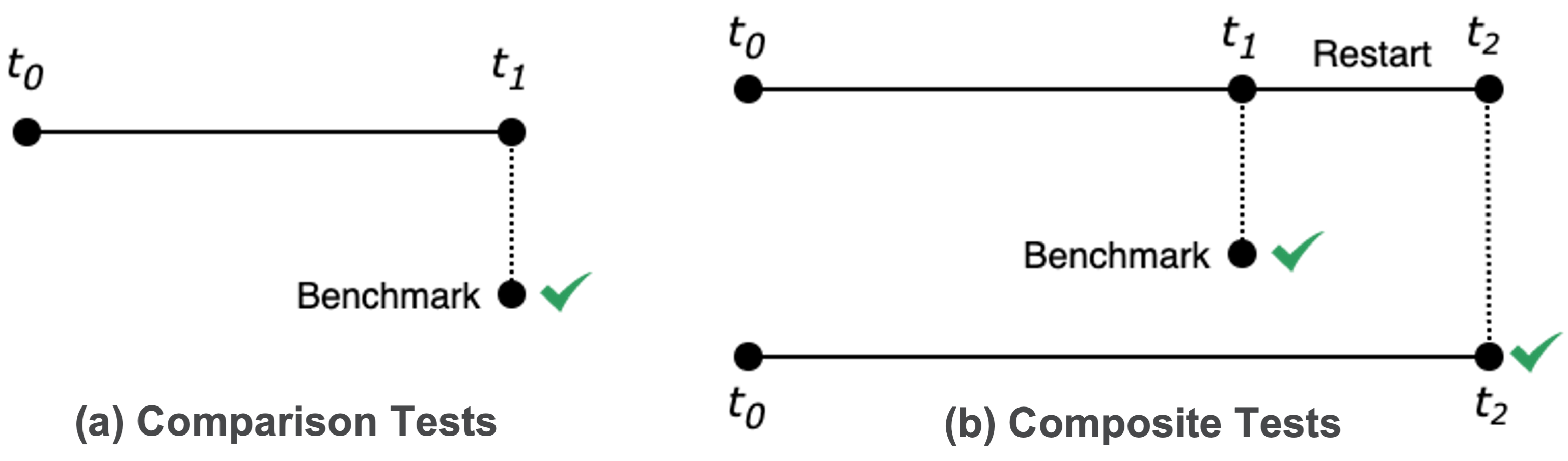}
\caption{Types of regression test requirements for simulation software instruments.}
\label{fig:test-types}
\end{center}
\end{figure}

\flashx's test specifications are essentially a collection of
options to be used to configure the applications for testing
purposes along with an accompanying {\em .par} file. The options are
specified during execution of \flashx's {\em setup} command, which also indicates which
parfile to use. We support unit tests and regression tests. As discussed above, our unit
tests do not strictly follow the definition as understood by the
broader community.  However, as expected, they do not rely on benchmarks.
Regression tests are
{\em comparison} or {\em composite} tests as shown in Fig \ref{fig:test-types}. Comparison tests are based on the
availability of an inspected and approved benchmark to compare the output
generated by the application instance run as a test of
correctness. Composite tests include two comparison tests, one in which
the application starts from initial conditions and runs to a time
$t_1$ and another in which the application starts from the checkpoint at
time $t_1$ and runs to a time $t_2$. Therefore, each composite test requires two
benchmarks.  Composite tests obviate the need for comparison tests on
the corresponding application setups.
\begin{figure}[h]
\begin{center}
\includegraphics[width=0.425\textwidth]{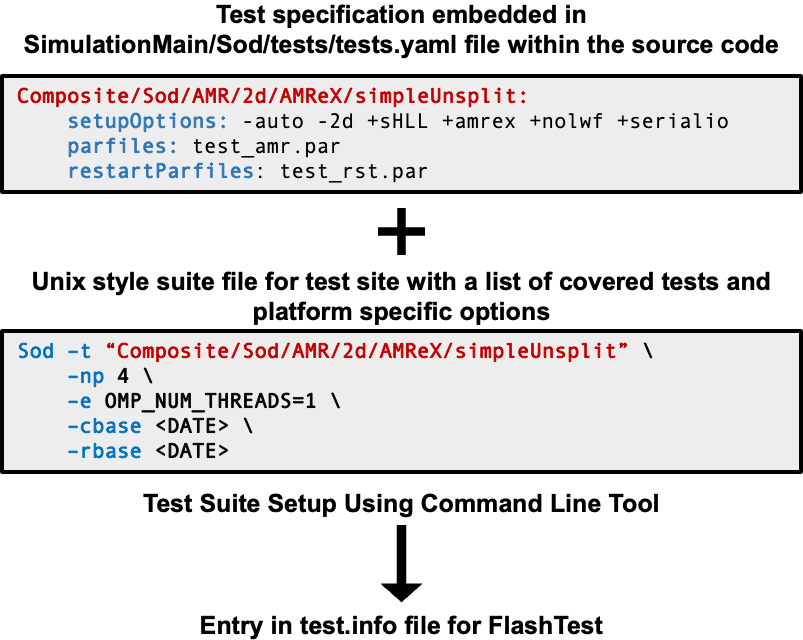}
\caption{Schematic of Yaml and Suite files that are combined to produce
a {\em test.info} file (see Section~\ref{testframework}) to run tests. The corresponding {\em test.info} is shown in Fig. \ref{fig:flashtest}, and the details related to the structure of the files are provided in Section \ref{sc:tooling}.}
\label{fig:suite-setup}
\end{center}
\end{figure}

To satisfy the first testing framework requirement, all the information needed to configure a set of application instances as
a set of related tests is encoded in a single {\em tests.yaml} file located under the
associated simulation's {\em tests} folder, along with the corresponding {\em
parfiles} and preprocessing scripts associated with individual tests. To
satisfy the second requirement of the testing framework, a single test suite dedicated
to a particular platform and software stack can be constructed outside the
code repository as a {\em suite} file by the Flash-X development team or, indeed,
by any Flash-X user.  The file contains an entry for each test to be included
in the test suite, and each line includes additional site-specific information
related to the test such as the location of
the benchmarks and the number of processors to be used to run the test. Within the context of \flashx, a {\em site} can be a personal workstation or a multinode cluster where users/developers wish to compile and run simulations. Each {\em site} is described by a {\em Makefile.h} that specifies the compiler, install location of external dependencies and various compile/link flags. Fig. \ref{fig:suite-setup} shows an example design of the test suite, which is described in the following subsections.

\subsection{Framework}
\label{testframework}
\begin{figure}[h]
\begin{center}
\includegraphics[width=0.425\textwidth]{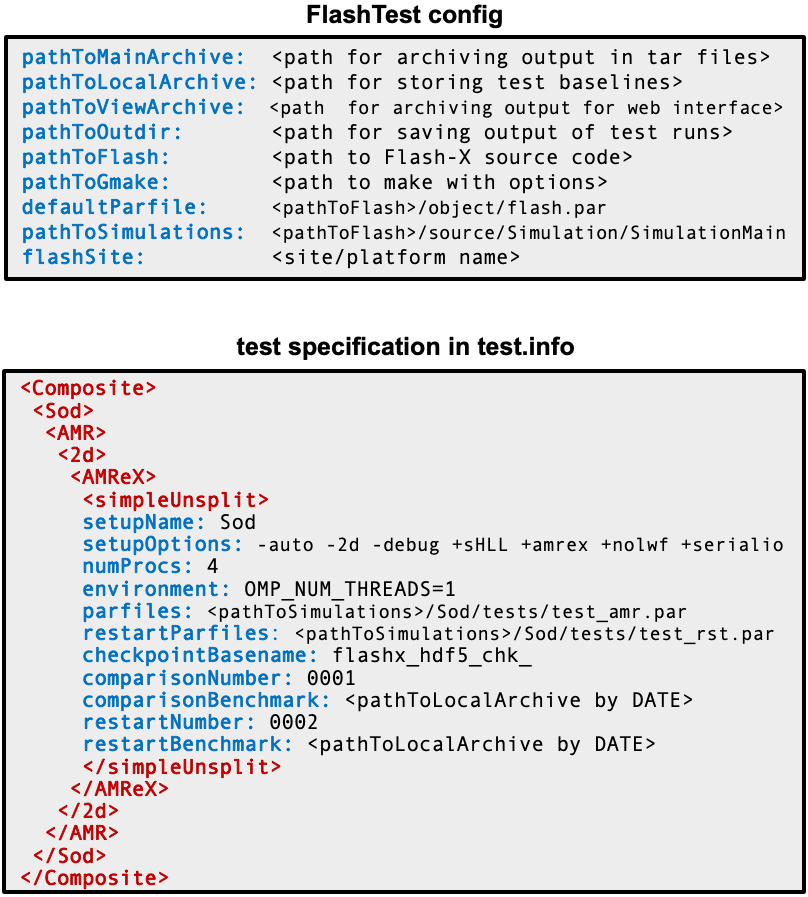}
\caption{Example of configuration files for \flashtest. The {\em config} file is created through the command-line invocation and {\em test.info} is generated by combining the {\em tests.yaml} and {\em suite} files, as shown in Fig. \ref{fig:suite-setup}. Details are provided in Section \ref{sc:tooling}.}
\label{fig:flashtest}
\end{center}
\end{figure}

\flashtest reads complete test specifications that include
site-specific details from an XML-format file called {\em test.info}.
{\flashtest also relies on other files to configure the execution
of tests at a {\em site}. Details related to resource requirements, location of 
benchmarks, \textit{etc.}, are stored in a set of configuration files that are isolated
from the source code and are under the control of the test executor.} See Fig.
\ref{fig:flashtest} for an example of {\em config} and {\em test.info}
files.

\begin{figure}[h]
\begin{center}
\includegraphics[width=0.35\textwidth]{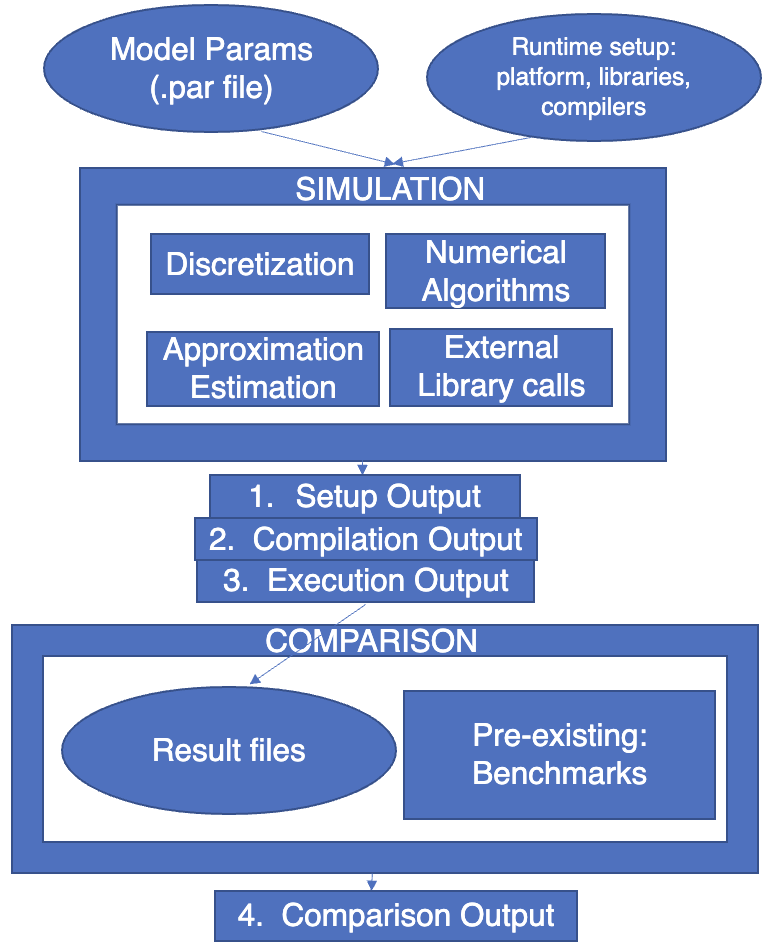}
\caption{Errors can be detected in the output files (1-4) in a simulation with \flashtest.}
\label{fig:workflow}
\end{center}
\end{figure}

Originally, {\em test.info} files  were written manually and independently for each site. This methodology was not particularly challenging when
development was confined largely within one institution with
infrequent external contributions.  Reliance was placed on internal
testing servers for regular testing needs. If at any point test specifications changed on
any server, the changes were manually conveyed and resolved everywhere
for consistency.  External collaborators
either did not run the test suite or independently configured their own
infrastructure. While \flashx initially followed the same methodology
in its early development stages, we quickly realized that manual configuration and
hidden internal test specifications are not compatible with the open
development model we have now adopted for the code.

\begin{figure}[h]
\begin{center}
\includegraphics[width=0.425\textwidth]{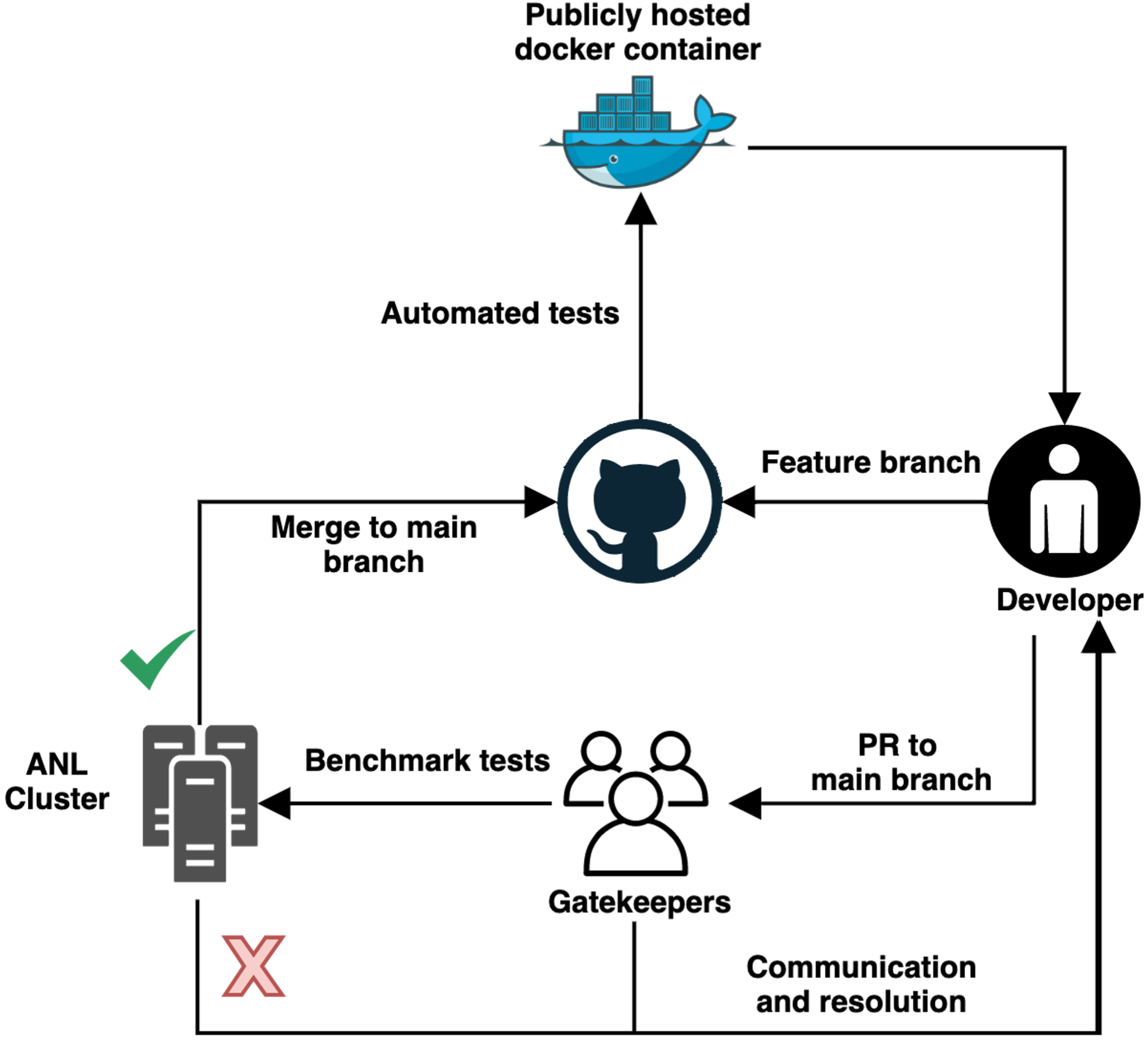}
\caption{Testing workflow for Flash-X: Individual developers require a streamlined process for setting up automated tests when developing new features, and gatekeepers require well-defined testing protocols to ensure quality of scientific results.}
\label{fig:framework-reqs}
\end{center}
\end{figure}

\flashx has brought together developers from different scientific
domains who require a more permissive testing infrastructure to manage
modifications to test specifications at multiple testing {\em sites}.
They need means to systematically setup configuration files on
new {\em sites}, and a solution that avoids manual copy/paste tasks. As a result, the \flashx testing
infrastructure needed an evolution to balance the trade-offs between
development and gatekeeping requirements (see Fig \ref{fig:framework-reqs}). Therefore, we built a tool that generates {\em test.info} from the
information contained in the {\em tests.yaml} and {\em suite} files described
above. The innovation of embedding an
individual test specification within the \flashx code repository where
quality review of test specifications can be done along with source pull
requests is also a direct consequence of this balancing act. 
We have also created a command-line toolkit that acts as a wrapper over
\flashtest to (1) organize and implement uniform testing practices across multiple
teams/platforms, and (2) automate the configuration of individual test suites so that
developers can easily construct and customize test suites on their own
development platform to suite their own development needs.

This is important since the natural interdependence between various code
components and the number of different ways that code components can
be configured often require that a developer modifying one particular component
run multiple tests on their feature branch to confirm that modifications do not
break other components. Compressible hydrodynamic solvers in \flashx are good
examples where such needs arise. \flashx supports two interchangeable
solvers with a third one planned for the near future. They can be
configured with different number of state variables,
which can become huge if a nuclear or a chemical reaction network is
included. They support multiple coordinate systems such as Cartesian,
cylindrical, spherical, and polar, and they can use many different
reconstruction schemes. Any modification in how
one solver interacts with the rest of the code can
adversely affect the other. Therefore, a developer modifying either of
the two solvers will need to run roughly a dozen or more tests on
their feature branch regularly to ensure that nothing is broken. The
toolkit makes it possible to configure a custom test suite on the
feature branch running on a local platform in a matter of minutes. 

\begin{figure}[h]
\begin{center}
\includegraphics[width=0.45\textwidth]{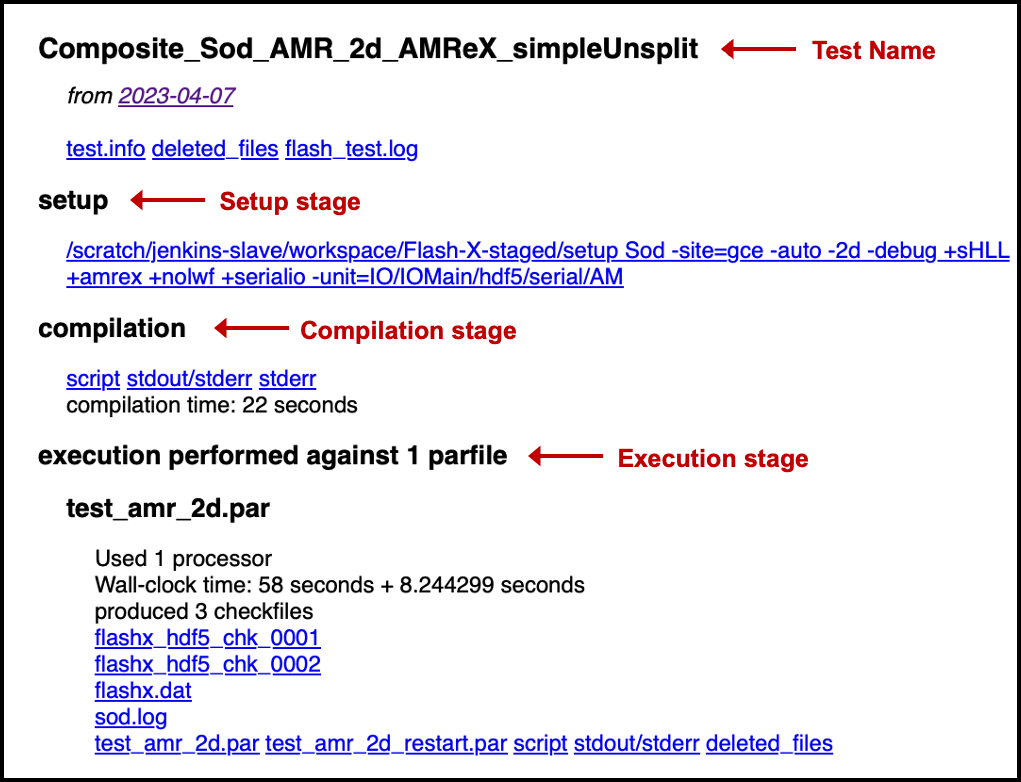}
\caption{Web interface for monitoring and viewing test results}
\label{fig:web-gui}
\end{center}
\end{figure}

\subsection{Stages of a Test}
\label{testingstages}
{Testing of an application instance in \flashx is performed in four different stages. Fig \ref{fig:workflow} outlines the workflow that \flashtest implements to assemble a test and identify errors during different stages as follows,
\begin{enumerate}
    \item {\bf Setup stage:} During this stage \flashx's internal \setuptool
    assembles an instance of the application for a test based on the {\em
    setupOptions} supplied in the {\em tests.yaml} file. This stage checks the internal design and organization of the source code.
    \item {\bf Compilation stage:} Successful completion of the setup stage triggers compilation of the application. The compilation stage depends on the {\em site}-specific {\em Makefile.h} and checks for syntax errors
in the source code, issues with compilation flags, and inability to link to external libraries.
    \item {\bf Execution stage:} The execution stage tests the runtime performance of the code using potentially multiple MPI processes and confirms the successful execution of unit tests. 
    \item {\bf Comparison stage:} This stage compares the test execution output with a verified benchmark and reports success/failure based on spatial errors between two data sets. Error checks are performed using the Serial Flash Output Comparison Utility (SFOCU), which is a custom-designed command-line tool to compare \flashx outputs.
\end{enumerate}
}

The results of the above stages can be viewed directly from the command
line or using the FlashTestView web interface shown in Fig. \ref{fig:web-gui}.  The interface presents all the test results with an interactive interface, making it easy to trace errors.


\begin{figure}[h]
\begin{center}
\includegraphics[width=0.425\textwidth]{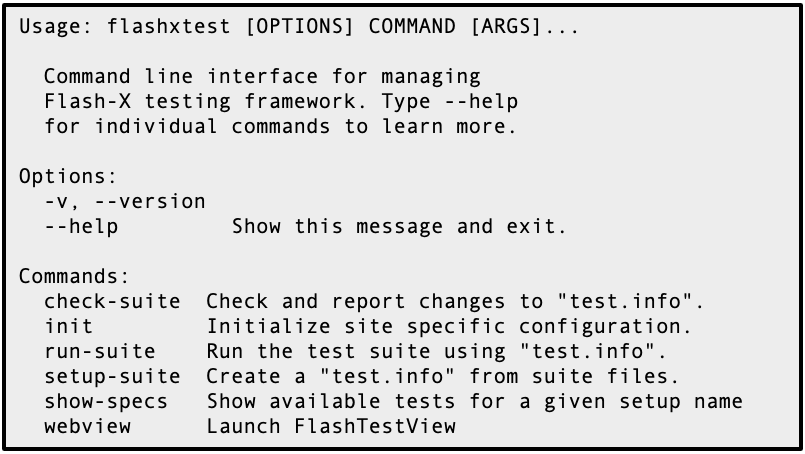}
\caption{command-line interface for \flashx's testing infrastructure}
\label{fig:cli}
\end{center}
\end{figure}

\section{Tooling}
\label{sc:tooling}

Fig \ref{fig:cli} provides an overview of the command-line toolkit {\em
flashxtest}, which wraps \flashtest and is distributed as a Python package.
Installation is managed using {\em pip}, which enables systematic update and
version tracking. Developers can use this tool to configure site-specific
configuration files and manage their local testing infrastructure.  

As described in section \ref{testcomps} the toolkit follows \flashx's
design and organization philosophy to embed simulation-specific test
configurations within their respective
{\em tests} folders which contain relevant {\em parfiles} and {\em yaml} files. 
Developers can modify these files to add or update tests
that accompany a change in source code. Simulation-specific
{\em tests.yaml} files serve as a record of existing tests that can be
used for different purposes.


A typical process of setting up the testing infrastructure using {\em
flashxtest} is divided into three steps,

\subsection{{Initialization}} 

During this step, a {\em config} file similar to that in Fig.
\ref{fig:flashtest} is created to specify the location of the \flashx source
code, {\em site} name, path to the archive and output directories to save test results, and compilation options that are not included in the {\em site} directory. This is achieved by running {\em flashxtest init} with relevant options \cite{flashx-test}. The {\em config} file can be ephemeral, recreated during the invocation of every test run, or saved and version controlled depending on the {\em site}-specific test management.

\subsection{{Setting up a Suite}} 

This step creates a {\em test.info} derived from a
given {\em suite} file. Each entry in a {\em suite} file consists of commands in the style of \flashx's \setuptool
that specifies which test to run for a setup-name along with options for
specifying the number of MPI processes, environment variables, and date of
comparison ({\em cbase}) and restart ({\em rbase}) benchmarks following the
style of \flashx's \setuptool. The {\em flashxtest setup-suite} command creates
{\em test.info} by combining information from {\em tests.yaml} and {\em suite}
as shown in Fig \ref{fig:suite-setup}. The {\em suite} files are typically managed
by individuals/gatekeepers who own them, and therefore are not required to be exposed to
the community. In the context of scientific software development, we refer
to {\em gatekeeping} as a quality assurance practice that maintains an authoritative
testing infrastructure under the guidance of the code management team and is managed by a designated group of {\em gatekeepers}.

In some cases, it may become necessary to create a new {\em test.info}
file using an existing {\em test.info} as a {\em seed} file. The command {\em
setup-suite} allows this by combining information from
the {\em seed.info}, {\em suite}, and {\em tests.yaml} and determining priorities
between conflicting attributes. This situation typically arises when {\em
seed.info} undergoes an update through the web interface, where users/developers can
update the benchmark information that leads to changes in values for the {\em
comparisonBenchmark} and {\em restartBenchmark} shown in Fig.
\ref{fig:flashtest}. The updated information in the {\em seed.info} file should take
precedence over the entries in the {\em suite} and {\em tests.yaml} files, and communicated
to developers and gatekeepers.

Changes in {\em yaml} files are discussed during pull requests, at
which point the {\em site} specific {\em test.info} files should be updated. Changes can be manual or automated, depending on the level of control desired by
developers/gatekeepers. For example, the process of accepting updates to {\em
yaml} files is automated for our community testing {\em site} provided by the
University of Tennessee.  However, for the authoritative testing {\em site} at
Argonne National Laboratory, updating {\em test.info} requires manual intervention after careful deliberation of differences compared to existing {\em seed.info}. Individual developers can operate in either mode.

\subsection{{Running a Suite}}

The final step is the {\em flashtext run-suite} command which runs
\flashtest using the {\em config} and {\em test.info}
files and reports the success or failure of each test. Results can be inspected through the web interface if it is installed, or through the logfile generated during the run. The results are also archived to appropriate locations during this step.



\section{TestSuite}




For a component-based scientific code like \flashx, ensuring adequate
coverage in testing is very non-trivial. Different permutations
and combinations of the code components result in different
application instances needed by different science
domains that use \flashx \cite{DubeyIJHPCA18, DubeyIJHPCA19}. The application setups that are part of the {\em Simulation} unit are primarily designed to run simulations. The first
challenge in devising a test is to determine parameters that will
allow the simulation to run quickly while still exercising all the
relevant features of the code that need to be exercised in order to confirm
correctness. For example, if an application uses a custom refinement
criterion for AMR, then the test must run long enough to refine at
least once. Therefore, comparison and composite tests typically require the participation
of domain scientists.

Once tests have been designed for all possible application instances,
the next challenge is to select which of them to run regularly to
ensure coverage. Running every available test is not practical,
because the test suite would take too long. Moreover, unless an
error is detected, running the entire set would waste
resources. Therefore, the nominal set of tests that are run for every
pull request are derived from the most complex applications. For
example, the test for {\em SNIa Double Detonation} application
exercises shock hydrodynamics, self gravity, nuclear burning with
multiple species, and specialized equation of state. If this test passes, then
there is no need to run any test that exercises a subset of these
components in a similar way. However, there must be finer tests that test the code
with simpler setups or at a lower level to help pinpoint the
cause of error if a test such as {\em SNIa Double Detonation} fails. The test
for {\em cellular detonation} exercises all the components that
{\em SNIa Double Detonation} does, except self gravity. If that
fails with the burning network used in {\em SNIa Double Detonation},
then the test can be run with a different burning network. If that also
fails, then one of several hydrodynamics only tests is used. This
methodology of repeatedly reducing the complexity of tests helps to
quickly narrow down the cause of failure.

\section{Conclusions}

In this paper, we provide a methodology for testing and verifying
complex scientific simulation software. The strategies we propose have been
shaped by the requirements that arise when maintaining a very complex instrument for
science and engineering workflows.

An open source development model requires implementation of gatekeeping practices to ensure research quality and reproducibility while balancing general community needs to streamline software development. Along with the design of our testing framework, we have also built a methodology to address the complexities of test coverage for highly composable systems like \flashx, and the trade-offs between optimizing the performance of a testing suite and creating a system to rapidly and efficiently trace sources of failures.

We believe that the design and workflow of our customized testing tool {\em flashxtest} meets the current needs of the \flashx community. We also believe that our solution can serve as a guide for designing testing frameworks for other scientific software systems where testing needs related to gatekeeping and general development are similar.

\section{Acknowledgements}
The submitted manuscript was created in part by UChicago Argonne, LLC,
operator of Argonne National Laboratory (“Argonne”). Argonne, a
U.S. Department of Energy Office of Science laboratory, is operated
under Contract No. DE-AC02-06CH11357. The U.S. Government retains for
itself, and others acting on its behalf, a paid-up nonexclusive,
irrevocable worldwide license in said article to reproduce, prepare
derivative works, distribute copies to the public, and perform
publicly and display publicly, by or on behalf of the Government.  The
Department of Energy will provide public access to these results of
federally sponsored research in accordance with the DOE Public Access
Plan. http://energy.gov/downloads/doe-public-access-plan. 

We also acknowledge the use of a large language model
(LLM) to correct grammatical errors in the text.

\bibliographystyle{abbrv}
\bibliography{biblio}

\begin{thebibliography}{10}

\bibitem{petsc-user-ref}
S.~Balay, S.~Abhyankar, M.~F. Adams, S.~Benson, J.~Brown, P.~Brune,
  K.~Buschelman, E.~Constantinescu, L.~Dalcin, A.~Dener, V.~Eijkhout,
  J.~Faibussowitsch, W.~D. Gropp, V.~Hapla, T.~Isaac, P.~Jolivet, D.~Karpeev,
  D.~Kaushik, M.~G. Knepley, F.~Kong, S.~Kruger, D.~A. May, L.~C. McInnes,
  R.~T. Mills, L.~Mitchell, T.~Munson, J.~E. Roman, K.~Rupp, P.~Sanan,
  J.~Sarich, B.~F. Smith, S.~Zampini, H.~Zhang, H.~Zhang, and J.~Zhang.
\newblock {PETSc/TAO Users Manual Revision 3.19}.
\newblock Technical Report ANL-21/39, Argonne National Laboratory, 2023.

\bibitem{bergerColella:1989}
M.~J. Berger and P.~Colella.
\newblock Local adaptive mesh refinement for shock hydrodynamics.
\newblock {\em Journal of Computational Physics}, 82(1):64--84, May 1989.

\bibitem{chawdhary2018immersed}
S.~Chawdhary, A.~Dhruv, A.~Dubey, and E.~Balaras.
\newblock Immersed boundary methods for fluid-structure interaction problems in
  two-phase flows with phase changes.
\newblock {\em Bulletin of the American Physical Society}, 63, 2018.

\bibitem{Dubey2009}
A.~Dubey, K.~Antypas, M.~Ganapathy, L.~Reid, K.~Riley, D.~Sheeler, A.~Siegel,
  and K.~Weide.
\newblock Extensible component based architecture for {FLASH}, a massively
  parallel, multiphysics simulation code.
\newblock {\em Parallel Computing}, 35:512--522, 2009.

\bibitem{DubeyPPAM2022}
A.~Dubey and T.~Klosterman.
\newblock Language agnostic approach for unification of implementation variants
  for different computing devices.
\newblock 2022.
\newblock Proceedings of PPAM 2022.

\bibitem{DubeyIJHPCA18}
A.~Dubey, P.~Tzeferacos, and D.~Lamb.
\newblock The dividends of investing in computational software design – a
  case study.
\newblock {\em International Journal of High Performance Computing
  Applications}, 2018.

\bibitem{dubey2018methodology}
A.~Dubey and H.~Wan.
\newblock Methodology for building granular testing in multicomponent
  scientific software.
\newblock In {\em 2018 IEEE/ACM 13th International Workshop on Software
  Engineering for Science (SE4Science)}, pages 9--15. IEEE, 2018.

\bibitem{dubey2015ongoing}
A.~Dubey, K.~Weide, D.~Lee, J.~Bachan, C.~Daley, S.~Olofin, N.~Taylor, P.~M.
  Rich, and L.~B. Reid.
\newblock Ongoing verification of a multiphysics community code: Flash.
\newblock {\em Software: Practice and Experience}, 45(2):233--244, 2015.

\bibitem{dubey2022flash}
A.~Dubey, K.~Weide, J.~O’Neal, A.~Dhruv, S.~Couch, J.~A. Harris,
  T.~Klosterman, R.~Jain, J.~Rudi, B.~Messer, et~al.
\newblock Flash-{X}: A multiphysics simulation software instrument.
\newblock {\em SoftwareX}, 19:101168, 2022.

\bibitem{falgout2000}
R.~Falgout and U.~Yang.
\newblock hypre: A library of high performance preconditioners.
\newblock {\em Computational Science-ICCS 2002}, pages 632--641, 2002.

\bibitem{DubeyIJHPCA19}
A.~Grannan, K.~Sood, B.~Norris, and A.~Dubey.
\newblock Understanding the landscape of scientific software used on
  high-performance computing platforms.
\newblock {\em accepted, International Journal of High Performance Computing
  Applications}, 2019.

\bibitem{harris2022exascale}
J.~A. Harris, R.~Chu, S.~M. Couch, A.~Dubey, E.~Endeve, A.~Georgiadou, R.~Jain,
  D.~Kasen, M.~P. Laiu, O.~B. Messer, et~al.
\newblock Exascale models of stellar explosions: Quintessential multi-physics
  simulation.
\newblock {\em The International Journal of High Performance Computing
  Applications}, 36(1):59--77, 2022.

\bibitem{RudiONeilWahibEtAl2021}
J.~Rudi, J.~O'Neal, M.~Wahib, and A.~Dubey.
\newblock {CodeFlow}: A code generation system for {Flash-X} orchestration
  runtime.
\newblock Technical Report ANL-21/17, Argonne National Laboratory, Lemont, IL,
  2021.

\bibitem{flashx-test}
{T}he~{F}lash {X}~{T}eam.
\newblock {F}lash-{X} {T}esting {T}ool.
\newblock
  \url{https://github.com/Flash-X/Flash-X-Test/tree/main/FlashXTest/example},
  2023.

\bibitem{flashx-userguide}
{T}he~{F}lash {X}~{T}eam.
\newblock {F}lash-{X} {U}serguide.
\newblock \url{https://flash-x.github.io/Flash-X-docs}, 2023.

\bibitem{Timmes1999}
F.~Timmes.
\newblock Integration of nuclear reaction networks.
\newblock {\em The Astrophysical Journal Supplement Series}, 124:241--263,
  1999.

\end{thebibliography}




\end{document}